\documentclass[10pt]{iopart}

\usepackage{iopams}  
\usepackage{cite} 
\usepackage{hyperref} 
\hypersetup{colorlinks=true, urlcolor=blue, linkcolor=blue, 
citecolor=blue} 
\usepackage{graphics} 
\usepackage{graphicx} 
\usepackage{mathrsfs}
\usepackage{amssymb}
\usepackage{bm}

\begin{document}

\title[Phase transitions and magnetization]{Phase transitions and magnetization of the mixed-spin Ising-Heisenberg double sawtooth frustrated ladder}

\author{Hamid Arian Zad}

\address{Young Researchers and Elite Club, Mashhad Branch, Islamic Azad University, Mashhad, Iran}
\eads{\mailto{\normalfont \color{blue} arianzad.hamid@mshdiau.ac.ir}}

\author{Nerses Ananikian}%
\address{Alikhanyan National Science Laboratory, Alikhanian Br. 2, 0036 Yerevan, Armenia}%
\eads{\mailto{\normalfont \color{blue} ananik@mail.yerphi.am}}
\vspace{10pt}
%\begin{indented}
%\item[]March 2015
%\end{indented}

\begin{abstract}
The mixed spin-(1,1/2) Ising-Heisenberg  double sawtooth ladder containing mixture of both spin-1 and spin-1/2 nodal atoms, and the spin-1/2 interstitial dimers is approximately solved by the transfer-matrix method. Here, we study in detail the ground-state phase diagrams, also influences of the bilinear exchange coupling on the rungs and cyclic four-spin exchange interaction in square plaquette of each block on the magnetization and magnetic susceptibility of the suggested ladder at low temperature. Such a double sawtooth ladder may be found in a Shastry-Sutherland Lattice-type. In spite of odd and even blocks spin ordering are different from each other, but due to the commutation relation between all different block Hamiltonians, phase diagrams,  magnetization behavior and thermodynamic properties of the model are the same for odd and even blocks. We show that at low temperature, both exchange couplings can change the quality and quantity of the magnetization plateaus versus the magnetic field changes. Specially, we find a new magnetization plateau  $\mathcal{M}/\mathcal{M}_{s}= 5/6$ for this model. Besides, we examine the magnetic susceptibility and specific heat of the model in detail. It is proven that behaviors of the magnetization and the magnetic susceptibility coincide at low temperature. The specific heat displays diverse temperature dependencies, which include a Schottky-type peak at a special temperature interval. We observe that with increase of the bilinear exchange coupling on the rungs, second peak temperature dependence grows.
\end{abstract}
%\submitto{\jpa}
%\maketitle
\section{Introduction} \label{sec:level1}
 Various versions of the spin chains has been encountered with a lot of attentions to display a wide variety of $T=0$ quantum phase transitions caused by changing the tunable parameters of the system$^,$s Hamiltonian such as external magnetic field and exchange couplings \cite{Ivanov,Kitaev,Dillenschneider,Sachdev,Werlang,Gu,Rojas2,RojasM,Ananikian,Abgaryan1,Abgaryan2,Strecka1,Valverde1}. A typical example is a  ladder of spins for which spins on the vertices have interaction along the legs by exchange coupling  $\mathcal{J}_\parallel$ and along the rungs via exchange coupling $\mathcal{J}_\perp$ \cite{Vuletic,Koga,Strecka,Toader,Okamoto,Muller}. Moreover, sawtooth chains as  another interesting solvable models have been studied from theoretical \cite{Arian1,Derzhko,Blundell,Bellucci,Buttner} and experimental \cite{Bacq,Ueland} point of view. 
 
 Geometrically frustrated Ising-like lattices in tetragonal shape have been theoretically and experimentally investigated. For instance, in Ref. \cite{Aronson} authors have synthesized single crystals of $Yb_2Pt_2Pb$  as a quasi-two dimensional system, where strong magnetic frustration may arise from the geometry of the underlying Shastry-Sutherland lattice-type (SSLT) and they reported experimental results of the magnetization, the specific heat, and magnetic susceptibility. They pointed out to existence of the Schottky-type peak in the specific heat and the deference in magnetic susceptibility behavior with respect to the temperature. In Ref. \cite{Dublenych} Y. Dublenych determine a complete and exact solution of the ground-state problem for the Ising model on the SSLT. He found a magnetization plateau at the one-third of the saturation value and  it was offered as the only possible fractional plateau in this model.
Recently, Linda Ye {\it et al.} studied on the Ising-type rare earth tetraborides $RB_4$ ($R =Er, Tm$), which is a tetragonal lattice topologically equivalent to the SSLT with visible magnetization plateaus \cite{Linda}. Furthermore, numerical calculations and related simulations for the magnetic susceptibility have been carried out. 
 
  In the last two decades, mixed spin models are one of the areas in the framework of the intensive research on the magnetism \cite{Hagiwara}.  We can refer to \cite{Ivanov1} as one of the beneficial references in which Ivanov presented a brief survey of the theoretical results in this hot area, also he investigated some basic quantum spin models of quasi one-dimensional quantum ferrimagnets with competing interactions, and compared them with each other.
 
On the theoretical side, mixed spin models have been constructed and discussed by several authors \cite{Abgaryan1,Ivanov1,Arian2,Verkholyak}. On the experimental side, besides the exactly solvable spin-1/2 models such as $CuPzN$ \cite{Jeong}, several real magnetic materials such as azurite $Cu_3(CO_3)_2(OH)_2$ and $Cu_3(TeO_3)_2Br_2$ can be described by Heisenberg spin models. Hence, motivated by the experimental observations, several authors have studied the case of antiferromagnetic mixed spin-(1,1/2) chain \cite{Ivanov1,Tonegawa}.  
T. Sakai and S. Yamamoto numerically showed a quantized magnetization plateau as a function of the field appearing at the one-third of the saturated magnetization for the mixed spin-(1,1/2) Heisenberg ferrimagnet with anisotropic exchange coupling in \cite{Sakai}.
Plateaus in magnetization curves as functions of the magnetic field and anisotropy parameter, also thermodynamic properties for mixed spin models have been investigated both numerically and analytically \cite{Hori}. In the most of studies, beside of the magnetization, some thermodynamic parameters of the mixed spin models have been investigated in detail. For example, specific heat which may exhibit Schottky-type peak at special temperature interval. Under the thermodynamic conditions, this peak might tend to a double-peak temperature dependence  \cite{Ivanov1,Verkholyak,Sakai,Karlova}. 

The diamond chain with Ising interaction have been exactly solved within the classical transfer-matrix technique in Ref. \cite{Valverde}. The magnetization plateaus for the mixed spin-(1,1/2) Ising diamond chain have been argued in \cite{Xin}.   In Ref. \cite{Ohanyan1} O. Rojas {\it et al.}  presented the analysis of the zero temperature ground-state phase diagrams for the mixed spin-(1,1/2) Ising-Heisenberg diamond chain and investigated the magnetization processes at low temperature, they obtained magnetization plateaus at $\mathcal{M}=1/5$ and $\mathcal{M}=3/5$ in the units of saturation magnetization. 

\begin{figure}
\begin{center}
\resizebox{0.8\textwidth}{!}{%
  \includegraphics{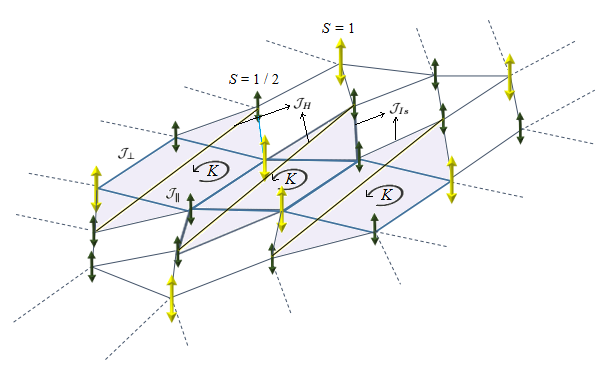}
}
%\vspace{5cm}       % Give the correct figure height in cm
\caption{Spin configuration for antiferromagnetic mixed-spin Ising-XXZ Heisenberg SSLT. Exchange couplings $\mathcal{J}_{\perp}$, $\mathcal{J}_{\parallel}$, $\mathcal{J}_{Is}$ and cyclic four-spin exchange coupling $K$ in square plaquette of the unit blocks (shaded areas) are shown. Note that each spin-1/2 has Heisenberg XXZ interaction with its fermionic next-next nearest neighbors. Shaded regions can be considered as a mixed-spin double sawtooth ladder in the  mixed-spin SSLT  with periodic boundary conditions when the number of blocks $\longrightarrow \infty$.}
\label{fig:figure1aa}  
\end{center}
\end{figure}

The approach we are offering in the present work is based on  the idea of using the mixed spin-(1,1/2) Ising-XXZ  double sawtooth ladder which can be found in a mixed-spin SSLT (figure \ref{fig:figure1aa}) as an approximately solvable spin model, and developing the transfer matrix formalism. The main goal of the present paper is to verify whether or not more intermediate magnetization plateaus can be detected at low temperature in a magnetization process of the  mixed spin-(1,1/2) Ising-XXZ double sawtooth ladder. Also, comparing the thermodynamic parameters of the model with each other is another significant intention of the paper.

This paper is organized as follow: in the next section we define the mixed spin double sawtooth ladder. Section \ref{QPT} deals with the most interesting results obtained for the ground-state phase diagrams at zero temperature. Further in section \ref{TMT}, we present the thermodynamic solution of the model using the enhanced transfer-matrix formalism. Indeed we show that, how one can investigate this spin model by using an approximate procedure. In section \ref{TM}, we have numerically discussed the magnetization, the magnetic susceptibility and the specific heat of the model at low temperature. Finally in section \ref{conclusion}, we summarize our results and draw the conclusions.
\section{Model}\label{Model}
\begin{figure}
\begin{center}
\resizebox{0.6\textwidth}{!}{%
  \includegraphics{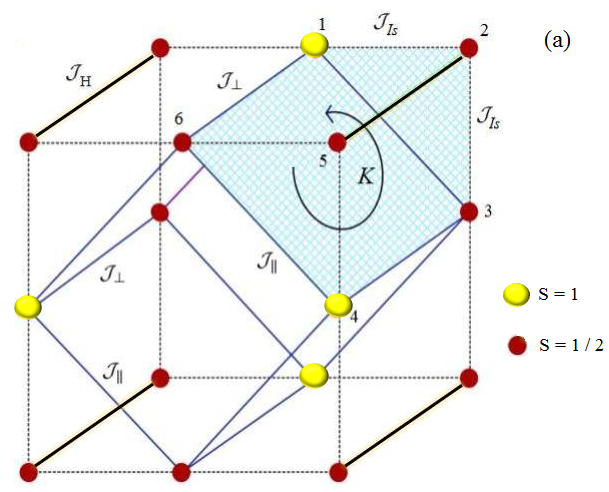}
}
\resizebox{0.4\textwidth}{!}{%
  \includegraphics{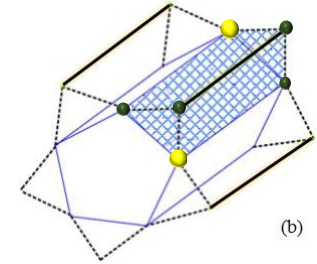}
}
\resizebox{0.4\textwidth}{!}{%
  \includegraphics{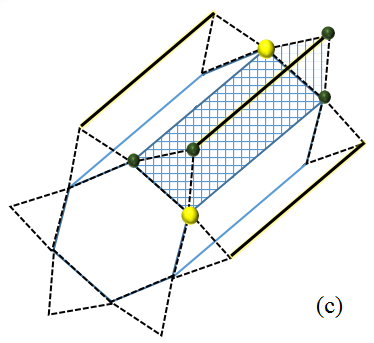}
}
%\vspace{5cm}       % Give the correct figure height in cm
\caption{Mixed-spin Ising-XXZ Heisenberg  double sawtooth ladder with geometric frustration for (a) $N=16$, (b) $N=20$ and  (c) $N=24$.}
\label{fig:figure1}  
\end{center}
\end{figure}

Recently, we investigated the Ising-XXZ Heisenberg double sawtooth ladder consist of half-spins in Ref. \cite{Arian1}. Phase transition and some thermodynamic parameters such as the heat capacity, the magnetization and magnetic susceptibility have been numerically investigated in detail.
In this work, we study the Ising-XXZ double sawtooth ladder with mixed nodal Ising spins including both spin-1/2 and spin-1 on the legs and interstitial dimer Heisenberg half-spins, in the presence of an external magnetic field. The disordered spin ladder of 16-spins with periodic boundary conditions is schematically illustrated in figure \ref{fig:figure1}(a). The number of spins in the ladder are selected even and our method is used for $N\geq 8$.
 To introduce a chain with $N>8$, the number of spins will grow as $N+4$, i.e., $N\in\{8,12,16,20,24\cdots\}$ (figures  \ref{fig:figure1}(b) and \ref{fig:figure1}(c) are related to $N=20$ and $N=24$, respectively). 
The Hamiltonian of the antiferromagnetic mixed spin double sawtooth ladder under the periodic boundary conditions is given by the following formula
\begin{equation}\label{SLhamiltonian}
\begin{array}{lcl}
H_{SL}=\\
\sum\limits_{i=1}^M\mathcal{J}_H {\boldsymbol\sigma}_{i,2}\cdot{\boldsymbol\sigma}_{i,5}+
\sum\limits_{i=1}^M\mathcal{J}_{\parallel} \big({J}_{i,1}^z{\sigma}_{i,3}^z+{J}_{i,4}^z{\sigma}_{i,6}^z\big)+ \\
\sum\limits_{i=1}^M\mathcal{J}_{Is} \big({J}_{i,1}^z{\sigma}_{i,2}^z+{\sigma}_{i,2}^z{\sigma}_{i,3}^z+{J}_{i,4}^z{\sigma}_{i,5}^z+{\sigma}_{i,5}
{\sigma}_{i,6}^z\big)+\\
\sum\limits_{i=1}^M\frac{\mathcal{J}_{\perp}}{2}\big({J}_{i,1}^z{\sigma}_{i,6}^z+{\sigma}_{i,3}{J}_{i,4}^z\big) +
K\sum\limits_{\langle 1346\rangle i}^{even}P_{1346}^{i\circlearrowleft}-\\
\sum\limits_{i=1}^{4M}\frac{{B}^{\prime z}}{2}\big({J}_{i,1}^z+{\sigma}_{i,3}^z+{J}_{i,4}^z+{\sigma}_{i,6}^z\big)-\sum\limits_{i=1}^{4M}{B}^{\prime\prime z}\big({\sigma}_{i,2}^z+{\sigma}_{i,5}^z\big),
\end{array}
\end{equation}
 where $M$ is the number of blocks (shaded regions in figure \ref{fig:figure1}), and $\mathcal{J}_H$ is the exchange coupling between half-spins of the interstitial Heisenberg dimer on each block and 
 \begin{equation}\label{Ringoperator}
 P^{\circlearrowleft}
\left(
\begin{array}{cc}
{S}_{i,1}^z & {s}_{i,3}^z \\
{s}_{i,6}^z & {S}_{i,4}^z \\
\end{array} \right) 
 =
 \left(
 \begin{array}{cc}
{s}_{i,3}^z & {S}_{i,4}^z \\
{S}_{i,1}^z & {s}_{i,6}^z \\
 \end{array}
 \right).
\end{equation}
 Here,  $2{\bf s}={\boldsymbol\sigma}=\lbrace {\sigma}^x, {\sigma}^y, {\sigma}^z \rbrace$ are the spin-1/2 operators (with $\hbar=1$) and
  \begin{equation}
\begin{array}{lcl}
 \sqrt{2}{S^x} = {J^x}=\left(
\begin{array}{ccc}
0 & 1 & 0\\
1 & 0 & 1\\
0 & 1 & 0 \\
\end{array} \right), 
\sqrt{2}{S^y} ={J^y}=\left(
\begin{array}{ccc}
0 & -i & 0\\
i & 0 & -i\\
0 & i & 0 \\
\end{array} \right),\\
{S^z} ={J^z}=\left(
\begin{array}{ccc}
1 & 0 & 0\\
0 & 0 & 0\\
0 & 0 & -1 \\
\end{array} \right),
\end{array}
\end{equation}
 where ${ S}^{x,y,z}$ are spin-1 operators, ${B}^{\prime z}$ and ${B}^{\prime\prime z}$ are applied homogeneous magnetic fields considered  in the $z$-direction.  $\mathcal{J}_{\perp}$ and $\mathcal{J}_{\parallel}$ are the bilinear exchange couplings on the rungs and along the legs of the block$^,$s plaquette, respectively. $K$ is the coupling of the cyclic four-spin permutation operator per plaquette, and $\mathcal{J}_{Is}$ is the Ising coupling between the spins on the legs of the block$^,$s plaquette and two spins of the interstitial Heisenberg dimer. 
Note that here, all of introduced parameters are considered dimensionless.

The Heisenberg part of the Hamiltonian $H_{SL}$ is introduced as 
\begin{equation}\label{HamiltonianT}
\begin{array}{lcl}
{\boldsymbol\sigma}_{i,2}\cdot{\boldsymbol\sigma}_{i,5}=\mathcal{J}_{x} \big({\sigma}^{x}_{i,2}{\sigma}^{x}_{i,5}+{\sigma}^{y}_{i,2}{\sigma}^{y}_{i,5}\big)+\Delta{\sigma}^{z}_{i,2}{\sigma}^{z}_{i,5}.\\
%\textbf{S}_{i,1}\cdot\textbf{s}_{i,3}+\textbf{S}_{i,4}\cdot\textbf{s}_{i,6}={S}^{z}_{i,1}{s}^{z}_{i,3}+{S}^{z}_{i,4}{s}^{z}_{i,6},\\
%\textbf{S}_{i,1}\cdot\textbf{s}_{i,2}+\textbf{s}_{i,2}\cdot\textbf{s}_{i,3}+\textbf{S}_{i,4}\cdot\textbf{s}_{i,5}+\textbf{s}_{i,5}\cdot\textbf{s}_{i,6}= \\
%{S}^{z}_{i,1}{s}^{z}_{i,2}+{s}^{z}_{i,2}{s}^{z}_{i,3}+{S}^{z}_{i,4}{s}^{z}_{i,5}+{s}^{z}_{i,5}{s}^{z}_{i,6},\\
%\frac{\textbf{B}^{\prime}}{2}\big(\textbf{S}_{i,1}+\textbf{s}_{i,3}+\textbf{S}_{i,4}+\textbf{s}_{i,6}\big)=\\
%\frac{{B}^{z\prime}}{2}\big({S}^{z}_{i,1}+{s}^{z}_{i,3}+{S}^{z}_{i,4}+{s}^{z}_{i,6}\big),\\
%\textbf{B}^{\prime\prime}\big(\textbf{s}_{i,2}+\textbf{s}_{i,5}\big)={B}^{z\prime\prime}\big({s}^{z}_{i,2}+{s}^{z}_{i,5}\big).
\end{array}
\end{equation}
The Ising-type cyclic four-spin permutation operator can be written as the product of three transposition operators $P_{1346}^{\circlearrowleft}=P_{13}^{\circlearrowleft}P_{14}^{\circlearrowleft}P_{16}^{\circlearrowleft}$ where $P_{13}^{\circlearrowleft}=1/2(1+{J}^{z}_{1}\sigma^{z}_{3})$, then we can obtain the following result which contains both bilinear and biquadratic terms of the spin-1/2 and spin-1 operators as 
\begin{equation}\label{FourC}
\begin{array}{lcl}
P_{1346}^{i\circlearrowleft}=\frac{1}{8}\big[1+{J}^{z}_{1}\sigma^{z}_{3}+{J}^{z}_{1}{J}^{z}_{4}+{J}^{z}_{1}\sigma^{z}_{6}+\\
({J}^{z}_{1}\sigma^{z}_{3})\cdot({J}^{z}_{1}{J}^{z}_{4})+({J}^{z}_{1}\sigma^{z}_{3})\cdot({J}^{z}_{1}\sigma^{z}_{6})+({J}^{z}_{1}{J}^{z}_{4})\cdot({J}^{z}_{1}\sigma^{z}_{6})+\\
({J}^{z}_{1}\sigma^{z}_{3})\cdot({J}^{z}_{1}{J}^{z}_{4})\cdot({J}^{z}_{1}\sigma^{z}_{6})\big].
\end{array}
\end{equation}
The $i$-th block Hamiltonian ${h}_{i}$ can be written as
\begin{equation}\label{HamiltonianT}
\begin{array}{lcl}
{h}_{i}=\big[\mathcal{J}_{x}\big({\sigma}^{x}_{i,2}{\sigma}^{x}_{i,5}+{\sigma}^{y}_{i,2}{\sigma}^{y}_{i,5}\big)+\Delta{\sigma}^{z}_{i,2}{\sigma}^{z}_{i,5}\big]\\
+\mathcal{J}_{\parallel}\big({J}^{z}_{i,1}{\sigma}^{z}_{i,3}+{J}^{z}_{i,4}{\sigma}^{z}_{i,6}\big)+\frac{\mathcal{J}_{\perp}}{2}\big({J}^{z}_{i,1}{\sigma}^{z}_{i,6}+{\sigma}^{z}_{i,3}{J}^{z}_{i,4}\big)\\
+\mathcal{J}_{Is}\big[{J}^{z}_{i,1}{\sigma}^{z}_{i,2}+{\sigma}^{z}_{i,2}{\sigma}^{z}_{i,3}+{J}^{z}_{i,4}{\sigma}^{z}_{i,5}+
{\sigma}_{i,5}{\sigma}^{z}_{i,6}\big]+K P_{1346}^{i\circlearrowleft}\\
-\frac{{B}^{\prime}_{z}}{2}\big({J}^{z}_{i,1}+{\sigma}^{z}_{i,3}+{J}^{z}_{i,4}+{\sigma}^{z}_{i,6}\big)-{B}^{\prime\prime}_{z}\big({\sigma}^{z}_{i,2}+{\sigma}^{z}_{i,5}\big).
\end{array}
\end{equation}
In this case, $\mathcal{J}>0$ is denoted antiferromagnetic exchange interactions and $\mathcal{J}<0$ ferromagnetic ones. The magnetic field $B^{\prime\prime}$ acts on the Heisenberg dimer spins and $B^{\prime}$ acts on the spins of the blocks plaquette with Ising-type interaction. Here, the case $B^{\prime}_{z}=2B^{\prime\prime}_z=2B$ is considered.
\section{Zero temperature phase diagrams}\label{QPT}
\begin{figure}
\begin{center}
\resizebox{0.6\textwidth}{!}{%
\includegraphics{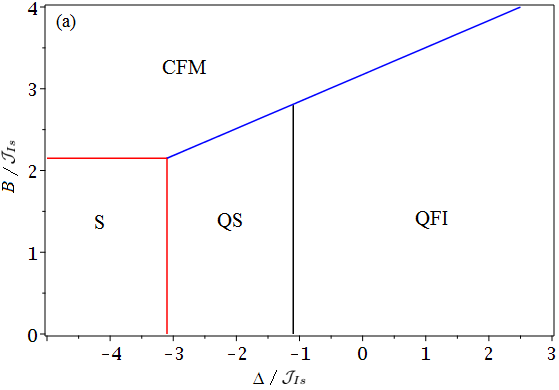}
}
\resizebox{0.6\textwidth}{!}{%
 \includegraphics{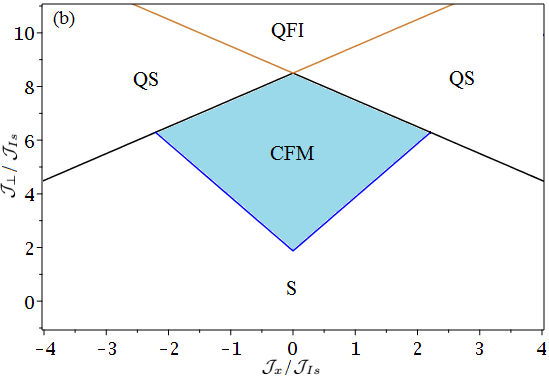}
 }
\caption{The ground-state phase diagrams of the mixed spin-(1,1/2) Ising-XXZ Heisenberg double sawtooth ladder (a) in the ($B/\mathcal{J}_{Is}-\Delta/\mathcal{J}_{Is}$) plane at fixed values of $\mathcal{J}_{x}=\mathcal{J}_{\parallel}=K=\mathcal{J}_{Is}$ and $\mathcal{J}_{\perp}=4\mathcal{J}_{Is}$, and (b) in the ($\mathcal{J}_{\perp}/\mathcal{J}_{Is}-\mathcal{J}_{x}/\mathcal{J}_{Is}$) plane at fixed values of $\Delta=-\mathcal{J}_{Is}$, $\mathcal{J}_{\parallel}=K=\mathcal{J}_{Is}$ and $B=4\mathcal{J}_{Is}$.}
\label{fig:QPTBDelta}       % Give a unique label 
\end{center}
\end{figure}

We here explain the ground-state phase diagrams of the introduced ladder. In figure \ref{fig:QPTBDelta}(a),
we display the zero temperature phase diagram in the ($B/\mathcal{J}_{Is}-\Delta/\mathcal{J}_{Is}$) plane by assuming fixed values of the another parameters applied in the Hamiltonian. Classical ferromagnetic (CFM), quantum ferrimagnetic phase (QFI), frustrated state (S) and quantum frustrated state (QS) are obtained. Spin arrangements of the relevant phases can be presented by using the following tensor product of the eigenvectors
\begin{equation}\label{PhaseD}
\begin{array}{lcl}
\vert CFM\rangle=\displaystyle\prod_{i=1}^{M}\vert 1\uparrow 1\uparrow\rangle_i\otimes\vert\varphi_1\rangle_i,\\
\vert QFI\rangle=\displaystyle\prod_{i=1}^{M}\vert 1\downarrow 1\downarrow \rangle_i\otimes\vert\varphi_2\rangle_i,\\
\vert S \rangle=\displaystyle\prod_{i=1}^{M}\vert 0\downarrow 0\downarrow\rangle_i\otimes\vert\varphi_1\rangle_i,\\
\vert QS\rangle=\displaystyle\prod_{i=1}^{M}\vert -1\uparrow -1\uparrow\rangle_i\otimes\vert\varphi_3\rangle_i,\\
\end{array}
\end{equation}
where
\begin{equation}\label{PhaseD}
\begin{array}{lcl}
\vert \varphi_1\rangle_i=\vert\uparrow\uparrow\rangle_i,
\vert \varphi_2\rangle_i=\frac{1}{2}(\vert\uparrow\downarrow\rangle+\vert\downarrow\uparrow\rangle)_i,\\
\vert \varphi_3\rangle_i=\frac{1}{2}(\vert\uparrow\downarrow\rangle-\vert\downarrow\uparrow\rangle)_i,
\vert \varphi_4\rangle_i=\vert\downarrow\downarrow\rangle_i,\\
\end{array}
\end{equation}
are the eigenstates of the interstitial XXZ Heisenberg dimer such that $\sigma^z\vert\uparrow\rangle=\vert\uparrow\rangle$ and $\sigma^z\vert\downarrow\rangle=-\vert\downarrow\rangle$. States $\vert 1\rangle$, $\vert 0\rangle$ and $\vert -1\rangle$ are set up in the ${J}^z$ basis states.
The ground-state energies for fixed values of $\mathcal{J}_{x}=\mathcal{J}_{\parallel}=K=\mathcal{J}_{Is}$ and $\mathcal{J}_{\perp}=4\mathcal{J}_{Is}$ are given by
\begin{equation}\label{PF}
\begin{array}{lcl}
\mathcal{E}_{CFM}=\frac{\Delta}{\mathcal{J}_{Is}}+11-\frac{6B}{\mathcal{J}_{Is}},
\mathcal{E}_{QFI}=-{\frac{\Delta}{\mathcal{J}_{Is}}-8}, \\
\mathcal{E}_{S}=\frac{\Delta}{\mathcal{J}_{Is}}-\frac{15}{8},
\mathcal{E}_{QS}=-{\frac{\Delta}{\mathcal{J}_{Is}}-4}.
\end{array}
\end{equation}
The blue line in figure \ref{fig:QPTBDelta}(a) separates regions of CFM and QFI phases with the curve $\frac{B}{\mathcal{J}_{Is}}=\frac{1}{3}\frac{\Delta}{\mathcal{J}_{Is}}+\frac{19}{6}$. The phase boundary between the QS and QFI states is limited by the line $\Delta\approx -1.1\mathcal{J}_{Is}$, and the boundary between the S and QS states is limited by the line $\Delta\approx -3.1\mathcal{J}_{Is}$. Line $B\approx 2.15\mathcal{J}_{Is}$ depicts the phase boundary between the S and CFM states.

Figure \ref{fig:QPTBDelta}(b) illustrates the ground-state phase diagram in the  ($\mathcal{J}_{\perp}/\mathcal{J}_{Is}-\mathcal{J}_{x}/\mathcal{J}_{Is}$) plane where from figure \ref{fig:QPTBDelta}(a) we consider fixed values $\Delta=-\mathcal{J}_{Is}$ and $\mathcal{J}_{\parallel}=K=\mathcal{J}_{Is}$ near the boundary between QS and QFI phases and a formal fixed value of magnetic field $B=4\mathcal{J}_{Is}$ due to continue our calculations to achieve second peak in the specific heat diagram. We prove that for other fixed values of the magnetic field the second peak occurs under spacial conditions at which the boundary of CFM phase may change. Shaded region shows CFM phase which plays an important role to describe the behavior of thermodynamic parameters of the favorite model surrounding critical points at which quantum phase transitions occur.

\section{Solution within the transfer-matrix technique}\label{TMT}
The mixed spin-(1,1/2) Ising-XXZ  double sawtooth ladder can be solved through the transfer matrix technique. The interstitial Heisenberg dimer coupling can be expressed as
\begin{equation}\label{density matrices}
\big({\boldsymbol\sigma}_{i,2}\cdot{\boldsymbol\sigma}_{i,5}\big)_{\Delta,\mathcal{J}_{x}}= \left(
\begin{array}{cccc}
\frac{\Delta}{4} & 0 & 0 & 0 \\
0 & -\frac{\Delta}{4} & \frac{\mathcal{J}_x}{2} & 0 \\
0 &\frac{\mathcal{J}_x}{2} & -\frac{\Delta}{4}  & 0\\
 0 & 0 & 0 & \frac{\Delta}{4}
\end{array} \right).
\end{equation}
Due to the commutation relation between different block Hamiltonians, $[{h}_{i},{h}_{j}]=0$, the partition function of the ladder can be written in the form
\begin{equation}\label{PF}
\mathcal{Z}=Tr\Big[\displaystyle\prod_{i=1}^{M}\exp(-\beta h_{i})\Big],
\end{equation}
where $\beta=\frac{1}{k_{B}T}$, $k_{B}$ is the Boltzmann’s constant and $T$ is the absolute temperature.
We can consider the following matrix representation for $\exp(-\beta h_{i})$ in the qubit-qutrit standard basis of the eigenstates of the composite spin operators $\{ J_{i,1}^{z},\sigma_{i,6}^{z},\sigma_{i,3}^{z},J_{i,4}^{z} \} $ on the two consecutive rungs of the  plaquette in block $i$. The partition function $\mathcal{Z}$ can be introduced as
\begin{equation}\label{PF}
\begin{array}{lcl}
\mathcal{Z}=Tr\big[ \langle J_{1,1}^z\sigma_{1,6}^z\vert\mathcal{T}\vert \sigma_{1,3}^z J_{1,4}^z\rangle\langle \sigma_{2,1}^z
J_{2,6}^z\vert \mathcal{T}\vert J_{2,3}^z\sigma_{2,4}^z\rangle\\ 
\cdots\langle \sigma_{M,1}^z J_{M,6}^z\vert \mathcal{T}\vert J_{M,3}^z\sigma_{M,4}^z\rangle \big],
\end{array}
\end{equation}
where $\sigma_{i,j}^z=\pm1$ and $J_{i,j}^z=\pm1, 0$ and
\begin{equation}\label{TrM}
\begin{array}{lcl}
\mathcal{T}(i)=\langle J_{i,1}^z\sigma_{i,6}^z\vert\exp(-\beta h_{i})\vert \sigma_{i,3}^z J_{i,4}^z\rangle=\\
\sum\limits_{k=1}^4\exp\big[-\beta\mathcal{E}_k(J_{i,1}^z\sigma_{i,6}^z,\sigma_{i,3}^z J_{i,4}^z)\big].
\end{array}
\end{equation}
Four eigenvalues of the $i-$th block with Hamiltonian $h_{i}$ are
\begin{equation}\label{eigenvalues1}
\begin{array}{lcl}
\mathcal{E}_1(i)={\Delta}+\mathcal{J}_{Is}\big(J_{i,1}^z+\sigma_{i,3}^z+J_{i,4}^z+\sigma_{i,6}^z\big)+\Xi-2B,\\
\mathcal{E}_2(i)=\\ 
-{\Delta}+\Xi+\sqrt{\mathcal{J}_{Is}^2\big(J_{i,1}^z+\sigma_{i,3}^z-J_{i,4}^z-\sigma_{i,6}^z\big)^2+4\mathcal{J}_{x}^2},\\
\mathcal{E}_3(i)=\\ 
-{\Delta}+\Xi-\sqrt{\mathcal{J}_{Is}^2\big(J_{i,1}^z+\sigma_{i,3}^z-J_{i,4}^z-\sigma_{i,6}^z\big)^2+4\mathcal{J}_{x}^2},\\
\mathcal{E}_4(i)={\Delta}-\mathcal{J}_{Is}\big(J_{i,1}^z+\sigma_{i,3}^z+J_{i,4}^z+\sigma_{i,6}^z\big)+\Xi+2B,
\end{array}
\end{equation}
where 
\begin{equation}\label{Xi}
\begin{array}{lcl}
\Xi=\mathcal{J}_{\parallel}\big(J_{i,1}^{z}\sigma_{i,3}^{z}+J_{i,4}^{z}\sigma_{i,6}^{z}\big)+\frac{\mathcal{J}_{\perp}}{2}\big(J_{i,1}^{z}\sigma_{i,6}^{z}+\sigma_{i,3}^{z}J_{i,4}^{z}\big)+\\
\frac{K}{8}\big[J_{i,1}^{z3}\sigma_{i,3}^{z}J_{i,4}^{z}\sigma_{i,6}^{z}+J_{i,1}^{z2}\sigma_{i,3}^{z}J_{i,4}^{z}+
J_{i,1}^{z2}J_{i,4}^{z}\sigma_{i,6}^{z}+\\
J_{i,1}^{z2}\sigma_{i,3}^{z}\sigma_{i,6}^{z}+J_{i,1}^{z}J_{i,4}^{z}+J_{i,1}^{z}\sigma_{i,3}^{z}+J_{i,1}^{z}\sigma_{i,6}^{z}+1\\
-B\big(J_{i,1}^{z}+\sigma_{i,3}^{z}+J_{i,4}^{z}+\sigma_{i,6}^{z}\big)-2B\big].
\end{array}
\end{equation}
We can figure out the transfer matrix $\mathcal{T}(i)$ as
\begin{equation}\label{Tmat}
\mathcal{T}(i) = \left(
\begin{array}{cccccc}
 \mathcal{T}_{11} &  \mathcal{T}_{12} &  \mathcal{T}_{13} &  \mathcal{T}_{14} &  \mathcal{T}_{15} & \mathcal{T}_{16} \\
\mathcal{T}_{21} &  \mathcal{T}_{22} &  \mathcal{T}_{23} &  \mathcal{T}_{24} &  \mathcal{T}_{25} & \mathcal{T}_{26} \\
\mathcal{T}_{31} &  \mathcal{T}_{32} &  \mathcal{T}_{33} &  \mathcal{T}_{34} &  \mathcal{T}_{35} & \mathcal{T}_{36} \\
\mathcal{T}_{41} &  \mathcal{T}_{42} &  \mathcal{T}_{43} &  \mathcal{T}_{44} &  \mathcal{T}_{45} & \mathcal{T}_{46} \\
\mathcal{T}_{51} &  \mathcal{T}_{52} &  \mathcal{T}_{53} &  \mathcal{T}_{54} &  \mathcal{T}_{55} & \mathcal{T}_{56} \\
\mathcal{T}_{61} &  \mathcal{T}_{62} &  \mathcal{T}_{63} &  \mathcal{T}_{64} &  \mathcal{T}_{65} & \mathcal{T}_{66} \\
\end{array} \right).
\end{equation}
After straightforward calculations the elements of the transfer matrix are defined through eigenvalues (\ref{eigenvalues1}).

Since the mixed spin double sawtooth ladder is translational invariant and all of $h_i$ are  independent of the site $i$, equation (\ref{PF}) can be expressed as
\begin{equation}\label{ConvertedZ}
\mathcal{Z}=Tr\big[\mathcal{T}^M\big].
\end{equation}
The total partition function can also be expressed in terms of six eigenvalues of the transfer matrix $\mathcal{T}$
\begin{equation}\label{TotalZ}
\mathcal{Z}=\Lambda_1^{M}+\Lambda_2^{M}+\Lambda_3^{M}+\Lambda_4^{M}+\Lambda_5^{M}+\Lambda_6^{M}.
\end{equation}
 In the thermodynamic limit, it is sufficient to consider only the largest eigenvalue $\Lambda_{max}$ to calculate the partition function. 
 In Ref. \cite{Strecka} authors investigated an Ising-Heisenberg three-leg tube and obtained a $8\times 8$ transfer matrix including short components for which four eigenvalues are zero, while other four eigenvalues can be easily gained. But here, the situation (obtaining the largest eigenvalue of the transfer matrix) is much more complex than previous cases.
 
\begin{figure}
\begin{center}
\resizebox{0.8\textwidth}{!}{%
\includegraphics{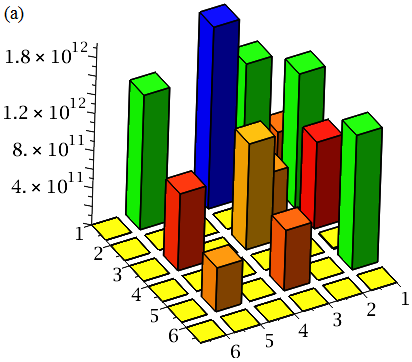}
\includegraphics{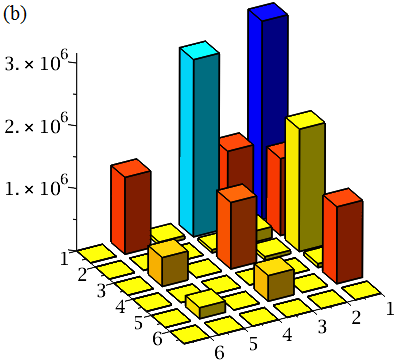}
}
\caption{Numerical matrix plots of the transfer matrix for fixed values of $\mathcal{J}_{\perp}/\mathcal{J}_{Is}=rand(1..8)$, $\mathcal{J}_{\parallel}/\mathcal{J}_{Is}=1$,$\mathcal{J}_{x}/\mathcal{J}_{Is}=rand(-2..2)$, $\Delta/\mathcal{J}_{Is}=rand(-1..6)$, $\beta\mathcal{J}_{Is}=rand(0..3)$ and $B/\mathcal{J}_{Is}=rand(0..8)$ (a) for 50-times randomly calculating and (b) over than 100-times. All minimum components (yellow regions) and the most effective components are unique for each-time randomly calculating.}
\label{fig:Matrixplot}
\end{center}
\end{figure}

Now, we are going to explain our approximate solution to extract the largest eigenvalue of the transfer matrix in detail. Since, we solely look for the largest eigenvalue of the transfer matrix, it is noteworthy that despite of the transfer matrix is $6\times 6$, but its elements are so long that obtaining the eigenvalues of  such a matrix is a big challenge for the accessible processors. We numerically investigated our favorite double sawtooth ladder and found that for the considered ranges of the Hamiltonian parameters, several components of the transfer matrix such as $\mathcal{T}_{66}$ are almost effectless and we can withdraw all of them. On the other hand, some components like $\mathcal{T}_{11}$, $\mathcal{T}_{13}$, $\mathcal{T}_{22}$ and other components out of yellow regions shown in figure \ref{fig:Matrixplot}  are the most effective components to determine the largest eigenvalue, which should be taken into account. Consequently, we reconstruct a new transfer matrix with less non-zero components. For another fixed values of the applied parameters in Hamiltonian, the elements of the transfer matrix are changed but according to our numerical investigations (randomly calculating over than 100-times) at low temperature, we realized that the most effective components of the transfer matrix deal with its largest eigenvalue are the same.
Accordingly, the new transfer matrix can be characterized as
\begin{equation}\label{NewTmat}
\mathcal{T}^{\prime}(i) = \left(
\begin{array}{cccccc}
 \mathcal{T}_{11} &  \mathcal{T}_{12} &  \mathcal{T}_{13} &  0 &  \mathcal{T}_{15} & 0 \\
\mathcal{T}_{21} &  \mathcal{T}_{22} &  0 &  0 &  0 & 0 \\
\mathcal{T}_{31} &  0 &  \mathcal{T}_{33} & 0 &  \mathcal{T}_{35} & 0 \\
0 &  0 &  0 & 0 &  0 & 0 \\
\mathcal{T}_{51} &  0 &  \mathcal{T}_{53} &  0 &  \mathcal{T}_{55} & 0 \\
0 &  0 &  0 &  0 &  0 & 0 \\
\end{array} \right).
\end{equation}
 Thus, we can consider the largest eigenvalue of the new transfer matrix $\mathcal{T}^{\prime}(i)$ equivalent to the largest eigenvalue of the primary transfer matrix $\mathcal{T}(i)$ with high accuracy.
 
 Free energy Gibbs per block for infinitely long chain, when lonely the maximal eigenvalue survives is given by \cite{Paulinelli}
 \begin{equation}\label{FreeE}
 \begin{array}{lcl}
 f=f_0+f^{\prime}=
 2\mathcal{J}_x+\Delta-\frac{1}{\beta}\lim\limits_{M\rightarrow \infty}\ln\frac{1}{M}\mathcal{Z}=\\
 2\mathcal{J}_x+\Delta-\frac{1}{\beta}\ln\Lambda_{max},
 \end{array}
 \end{equation}
\section{Thermodaynamic parameters}\label{TM}
Let us examine thermodynamic parameters of the model as functions of the magnetic field, the bilinear exchange coupling on the rungs $\mathcal{J}_{\perp}$ and the cyclic four-spin exchange interaction $K$ at low temperature. Note that the quantum spin ladders are the simplest spin systems for which the cyclic four-spin exchange appears from the interaction between nodal spins of the square plaquettes. This specific spin interaction  has been  theoretically and experimentally analyzed in the various spin models and some of its valuable applications (such as explaining the neutron-scattering) have been demonstrated \cite{Ivanov,Capponi,Roger1,Hakobyan}. The magnetization, the magnetic susceptibility and the specific heat can be obtained using the maximal eigenvalue of the transfer matrix  $\mathcal{T}^{\prime}(i)$ through the following formulae
\begin{equation}
\begin{array}{lcl}
%\mathcal{M}=\frac{1}{\beta}\frac{\partial\ln\Lambda_{max}}{\partial B},
  \mathcal{M}=-\frac{\partial f}{\partial B}, \chi=\frac{\partial\mathcal{M}}{\partial B}, \mathcal{C}=-k_B\beta\frac{\partial}{\partial\beta}\big(\beta^{2}\frac{\partial f}{\partial\beta}\big).
\end{array}
\end{equation}
It was shown that firstly, magnetization plateaus appear for a great number of spin models, and can be theoretically analyzed by the transfer matrix techniques as well as the exact diagonalization. Secondly, the spin gap existence in the spectrum of magnetic excitations creates plateaus at $\mathcal{M}/\mathcal{M}_{s}\neq 0$ in the external magnetic field.  On top of that, a number of spin-1/2 chains exhibit magnetization plateau at $\mathcal{M}/\mathcal{M}_{s}= 1/3$ and $\mathcal{M}/\mathcal{M}_{s}= 2/3$ which represents a massive phase \cite{Hida,Cabra}, while mixed spin-(1,1/2) chain \cite{Sakai} exhibits a plateau at $\mathcal{M}/\mathcal{M}_{s}= 1/2$, 
where $\mathcal{M}$ is the magnetization and $\mathcal{M}_{s}$ is called saturation magnetization. The zero magnetization plateau $\mathcal{M}/\mathcal{M}_{s}= 0$ for the Heisenberg spin ladders were obtained \cite{Arian1,Cabra}. Here, we investigate the magnetization plateaus for the considered mixed spin double sawtooth ladder. Among the variety of magnetization curves obtained for different sets of $B/\mathcal{J}_{Is}$ and $\mathcal{J}_{\perp}/\mathcal{J}_{Is}$, the most remarkable plateau is  $\mathcal{M}/\mathcal{M}_{s}= 1/3$ in the units of block.

\begin{figure}
\begin{center}
\resizebox{0.7\textwidth}{!}{%
\includegraphics{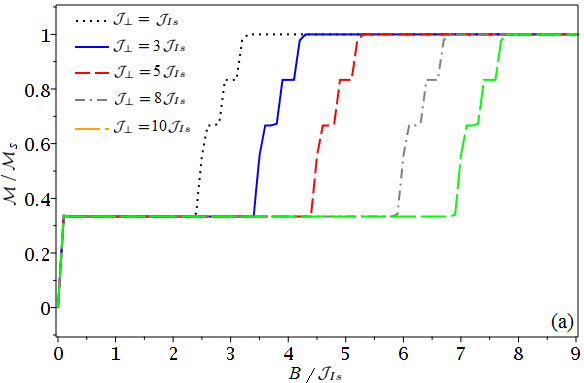}
}
\resizebox{0.7\textwidth}{!}{%
\includegraphics{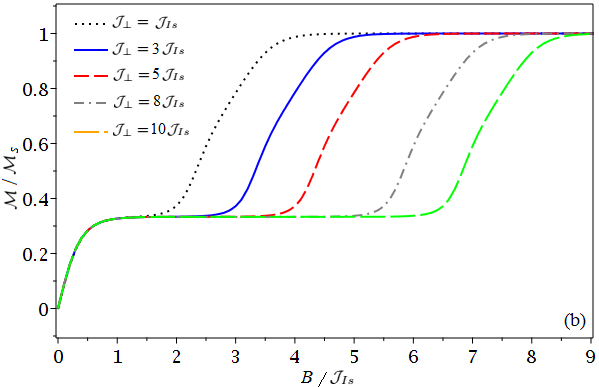}
}
\resizebox{0.7\textwidth}{!}{%
\includegraphics{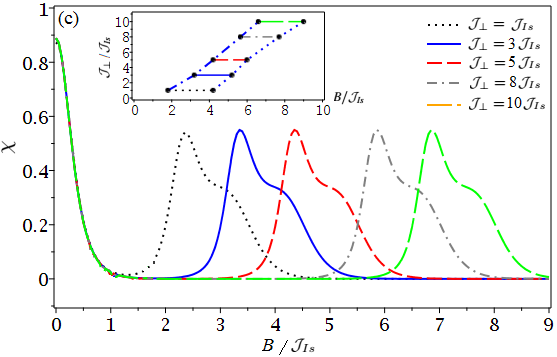} 
}

\caption{ The magnetization with respect to its saturation value as a function of $B/\mathcal{J}_{Is}$ for various fixed values of 
$\mathcal{J}_{\perp}/\mathcal{J}_{Is}$, where other parameters are taken as $\mathcal{J}_{\parallel}=K=\mathcal{J}_{x}=\mathcal{J}_{Is}$ and $\Delta=-\mathcal{J}_{Is}$ (a) at zero temperature, and (b) at low temperature $\beta=2\mathcal{J}_{Is}^{-1}$. (c) The magnetic susceptibility of the model per unit block as function of  $B/\mathcal{J}_{Is}$ for various fixed values of $\mathcal{J}_{\perp}/\mathcal{J}_{Is}$, where other parameters are taken as $\mathcal{J}_{\parallel}=K=\mathcal{J}_{x}=\mathcal{J}_{Is}$ and $\Delta=-\mathcal{J}_{Is}$, at low temperature $\beta=2\mathcal{J}_{Is}^{-1}$. The inset shows the critical points (solid circles) in the ($B/\mathcal{J}_{Is}$-$\mathcal{J}_{\perp}/\mathcal{J}_{Is}$) plane at which second double-peak of the magnetic susceptibility arises and vanishes, respectively, at low temperature.}
\label{fig:2DM}
\end{center}
\end{figure}

The difference in magnetization behavior with respect to the magnetic field is shown in figures \ref{fig:2DM}(a) and \ref{fig:2DM}(b). Figure  \ref{fig:2DM}(a) demonstrates the magnetization versus the magnetic field at zero temperature, while figure \ref{fig:2DM}(b) plots this function versus the magnetic field at low temperature.
It is quite obvious that the magnetization curve has a plateau at $\mathcal{M}/\mathcal{M}_{s}= 1/3$ for all  fixed values of 
$\mathcal{J}_{\perp}$ and $\Delta<0$ whether the system is considered at zero temperature or low temperature, on the other hand, this plateau gets wider as the ratio $\mathcal{J}_{\perp}/\mathcal{J}_{Is}$ strengthens. Interestingly at zero temperature, the magnetization exhibits two narrow plateaus at $\mathcal{M}/\mathcal{M}_{s}= 2/3$ and $\mathcal{M}/\mathcal{M}_{s}= 5/6$ which are not clear for the case where the system is considered at low temperature. A stronger evidence of two quantum phase transitions of the mixed spin-(1,1/2) Ising-XXZ double sawtooth ladder is provided by the magnetic susceptibility. By inspecting figure \ref{fig:2DM}(c), one can see that the magnetic susceptibility decreases upon increasing of the magnetic field from zero, until it vanishes at a critical magnetic field. Indeed, there is an identical rather steep decrease for the magnetic susceptibility function for all fixed values of $\mathcal{J}_{\perp}$ at weak magnetic field. When the magnetic field further strengthens, the magnetic susceptibility displays a double-peak for which second peak is much more smooth than first. The steep is taken in the interval
 $0<B/\mathcal{J}_{Is}\lesssim B_1/\mathcal{J}_{Is}$ and the double-peak appears in the interval
 $\frac{(B_1/\mathcal{J}_{Is}+\mathcal{J}_{\perp}/\mathcal{J}_{Is})}{2}+1<B/\mathcal{J}_{Is}\lesssim \frac{(B_1/\mathcal{J}_{Is}+\mathcal{J}_{\perp}/\mathcal{J}_{Is})}{2}+3$ (in this case $B_1\approx\mathcal{J}_{Is}$).
 
 Figure \ref{fig:2DM}(c) also shows  the dynamics of the magnetic susceptibility with respect to the exchange coupling
  $\mathcal{J}_{\perp}$. Namely, when  $\mathcal{J}_{\perp}$ increases the double-peak moves to the stronger magnetic fields. According to the phase diagram illustrated  in figure \ref{fig:QPTBDelta}(b), the steep shows S phase. The gap between the steep and the double-peak, where the magnetic susceptibility disappears, denotes CFM phase region. The double-peak coincides with QS phase and finally, the plateau next to the double-peak is in accordance with QF phase. Consequently, the CFM boundary gets wider upon increasing of $\mathcal{J}_{\perp}$.

\begin{figure}
\begin{center}
\resizebox{0.8\textwidth}{!}{%
\includegraphics{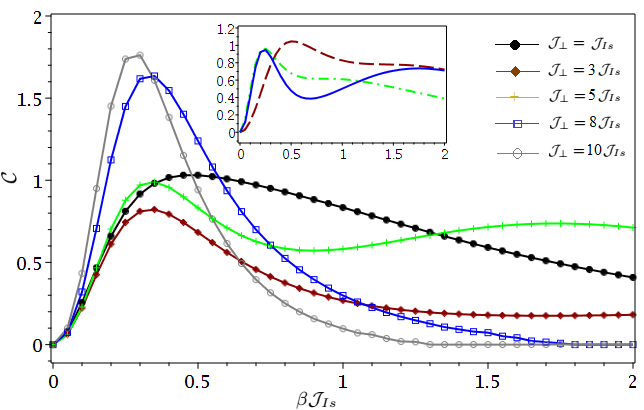}
}
\caption{Plot shows the specific heat of the mixed spin-(1,1/2) double sawtooth ladder as a function of the inverse temperature $\beta\mathcal{J}_{Is}$ for various fixed values of $\mathcal{J}_{\perp}$, where other parameters are taken as $\mathcal{J}_{\parallel}=K=\mathcal{J}_{x}=\mathcal{J}_{Is}$, $\Delta=-\mathcal{J}_{Is}$ and  $B=4\mathcal{J}_{Is}$. The inset shows the temperature dependence of the specific heat for several fixed values of the coupling constant  $\mathcal{J}_{\perp}$ at different magnetic fields. Brown dashed line depicts the specific heat changes at fixed $\mathcal{J}_{\perp}=3\mathcal{J}_{Is}$ and $B=3\mathcal{J}_{Is}$, green dashed-dot line at fixed $\mathcal{J}_{\perp}=5\mathcal{J}_{Is}$ and $B=6\mathcal{J}_{Is}$, and blue solid line at fixed $\mathcal{J}_{\perp}=8\mathcal{J}_{Is}$ and $B=5.5\mathcal{J}_{Is}$ (these fixed values have been taken up from the intervals containing critical magnetic fields).}
\label{fig:SH}
\end{center}
\end{figure}

Last but not least, let us examine the bilinear exchange couplings on the rungs  $\mathcal{J}_{\perp}$ and the magnetic field variations of the specific heat against the temperature. For this purpose, figure \ref{fig:SH} illustrates temperature dependence of the specific heat for the model under consideration with respect to the various fixed values of  $\mathcal{J}_{\perp}$.
As one can see, the specific heat exhibits a steep increase when the temperature gradually decreases. With further decreasing the temperature, the specific heat decreases and gradually tends to zero (black circle-line). There is solely one unique peak for the considered  fixed magnetic field $B=4\mathcal{J}_{Is}$ and $\mathcal{J}_{\perp}=\mathcal{J}_{Is}$. When $\mathcal{J}_{\perp}/\mathcal{J}_{Is}$  is lifted from 1, the single peak decreases and a second peak appears, which can be related to an abrupt $\mathcal{J}_{\perp}$-induced steep increase of the magnetization from plateau $\mathcal{M}/\mathcal{M}_{s}= 1/3$ toward saturation value. The second peak denotes CFM phase, where with further increase of $\mathcal{J}_{\perp}$ this peak disappears and simultaneously first peak becomes larger until reaches its maximum value ($\mathcal{C}\approx 1.8$). The highest value of the first peak may denote QFI phase. 

It is worthwhile to remark that by tuning the magnetic field the double-peak occurs for all fixed values of the coupling constant 
$\mathcal{J}_{\perp}$. In this regard, it could be quite interesting to answer the profound question; in which intervals of the magnetic field the double-peak may occurs? According to our investigations we found that the double-peak appears at the critical magnetic field $B_i\approx \frac{(B_1/\mathcal{J}_{Is}+\mathcal{J}_{\perp}/\mathcal{J}_{Is})}{2}+1$, and will remain until $B_f\approx\frac{(B_1/\mathcal{J}_{Is}+\mathcal{J}_{\perp}/\mathcal{J}_{Is})}{2}+3$ (inset of figure  \ref{fig:SH}). It is quite obvious from figure \ref{fig:2DM}(b) and its inset that in these intervals of the magnetic field, the double-peak occurs for the magnetic susceptibility function.
These intervals determine range of applicability of the  mixed spin-(1,1/2) Ising-XXZ double sawtooth  ladder for the purposes associated with ultra-cold atoms. 
The peaks in the magnetic susceptibility and the specific heat can be thus regarded as faithful indicators of the quantum critical points of the mixed spin-(1,1/2) double sawtooth  ladder.

\begin{figure}
\begin{center}
\resizebox{0.7\textwidth}{!}{%
\includegraphics{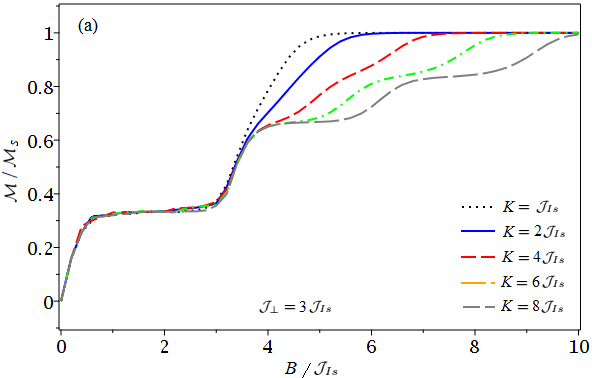}
}
\resizebox{0.7\textwidth}{!}{%
\includegraphics{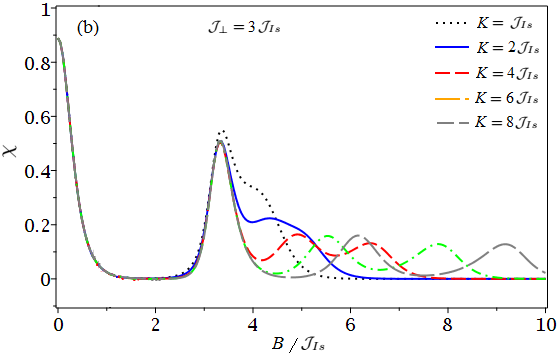}
}
\caption{(a) The magnetization and (b) the magnetic susceptibility as a function of $B/\mathcal{J}_{Is}$ for various fixed values of $K/\mathcal{J}_{Is}$ where other parameters are taken as $\mathcal{J}_{\parallel}=\mathcal{J}_{x}=\mathcal{J}_{Is}$, $\Delta=-\mathcal{J}_{Is}$ and $\mathcal{J}_{\perp}=3\mathcal{J}_{Is}$, at low temperature $\beta=2\mathcal{J}_{Is}^{-1}$.}
\label{fig:MPlat}
\end{center}
\end{figure}

Finally, let us turn back to the most spectacular magnetic field and exchange coupling $\mathcal{J}_{\perp}$ dependencies of the magnetization and magnetic susceptibility, which play an important role to describe quantum critical points of the investigated model. We examine the magnetization as function of the magnetic field for various fixed values of the cyclic four-spin exchange interaction $K$ at low temperature and arbitrary fixed $\mathcal{J}_{\perp}=3\mathcal{J}_{Is}$. We conceptually understand that $K$ has a substantial role to determine the number of magnetization plateaus. To support this statement, we have depicted in figure \ref{fig:MPlat}(a) typical cyclic four-spin exchange interaction variations of the magnetization against the magnetic field. One finds by inspection three separated plateaus 
$\mathcal{M}/\mathcal{M}_{s}= 1/3$, $\mathcal{M}/\mathcal{M}_{s}= 2/3$ and  $\mathcal{M}/\mathcal{M}_{s}= 5/6$ (a rare $K$-dependent plateau), which the last two plateaus gradually appear upon increasing the cyclic four-spin exchange interaction $K$. 

Figure \ref{fig:MPlat}(b) plots the magnetic susceptibility versus magnetic field for several fixed values of $K$ at low temperature. We find that with increase of $K$,  peaks of the mentioned double-peak are gradually separated from each other, which denotes second magnetization plateau $\mathcal{M}/\mathcal{M}_{s}= 2/3$ appearance. With further increase of $K$ another new peak grows and gets away from separated peaks, namely, magnetic field gaps will arise between all peaks. At low temperature, these gaps and all available magnetization plateaus coincide exactly.

\section{Conclusions}\label{conclusion}
In the present article, we have examined the ground state and thermodynamics of the mixed spin-(1,1/2) Ising-XXZ double sawtooth ladder including interstitial Heisenberg dimer half-spins connected to the leg sites on each block by taking advantage of the transfer-matrix technique. First of all, we have obtained exact results for the ground-state phase diagrams. Then, we have numerically investigated the transfer matrix and realized that the most of components are almost effectless to derive its largest eigenvalue, so they were eliminated and a new transfer matrix with less components was generalized. The new transfer matrix technique reported in this paper offers an approximate procedure to solve the model. This procedure provides the possibility of obtaining analytical expressions for the thermodynamic quantities such as the magnetization, the magnetic susceptibility and the specific heat with high accuracy. 

The magnetization have been examined in detail with respect to the magnetic field and the bilinear exchange couplings on the rungs at low temperature. Under this circumstances, we just observed the plateau $\mathcal{M}/\mathcal{M}_{s}= 1/3$ for various fixed values of  the bilinear exchange couplings on the rungs. On the other hand, we observed a plateau at $\mathcal{M}/\mathcal{M}_{s}= 2/3$ upon increasing the cyclic four-spin exchange interaction. The most interesting finding stemming from our study certainly represents a striking plateau  $\mathcal{M}/\mathcal{M}_{s}= 5/6$ upon further increase of the cyclic four-spin exchange interaction. This plateau appears rarely for spin models. Furthermore, we have investigated the magnetic susceptibility and compared it with the magnetization. We found that the magnetization behavior is in accordance with the magnetic susceptibility changes, where by increasing the cyclic four-spin exchange interaction, peaks of the field-dependent double-peak are gradually separated from each other and a magnetic field gap grows between them. As the cyclic four-spin exchange interaction further increases, an extra peak appears and get away from other peaks, namely, between any two consecutive peaks a distinguishable magnetic field gap arises. Consequently, magnetization plateaus and magnetic susceptibility gaps coincide at low temperature.

Finally, we have investigated the specific heat  with respect to the inverse temperature for several values of the bilinear exchange couplings on the rungs, where other parameters applied in the Hamiltonian were taken as fixed values. It is quite surprising that at low temperature we observed a Schottky-type peak. When  the bilinear exchange couplings on the rungs increases second peak arises, so under this condition, the specific heat has a double-peak. The temperature dependence of this function is almost similar to the specific heat depicted in Ref. \cite{Ivanov1}. Namely, when the temperature increases, the specific heat gradually increases and reaches its maximum at a special temperature interval. Here, this thermodynamic parameter have been more investigated numerically and remarkable outcomes have been obtained. For instance, we obtained critical intervals of the magnetic field at which the specific heat shows a double-peak for various fixed values of the  bilinear exchange couplings on the rungs. Indeed, we found that the range of these critical intervals are dependent on the fixed value of the  bilinear exchange couplings on the rungs. Interestingly, in these intervals the field-dependent double-peak of the magnetic susceptibility occurs for the corresponding fixed values of the bilinear exchange couplings on the rungs.

 Major applications of the considered mixed-spin double sawtooth  ladder in the real world are saving and transferring quantum information (both qbits and qtrits) through the model which surely can be useful in quantum information processing, optical lattices, spintronics and applied condensed matter physics.

We note that the precise method elaborated in the present paper is valid at low temperature and can be straightforwardly adapted to account for the another spin models with more complicate transfer matrix and different spin arrangement.

\section{Acknowledgments}
NA acknowledge by partly financial support of  the  MC-IRSES  no. 612707 (DIONICOS) under FP7-PEOPLE-2013 and  ICTP NT-04 grants.

%
% BibTeX users please use
% \bibliographystyle{}
% \bibliography{}
%\begin{thebibliography}{}
% \bibliographystyle{bib}
{
\end{document}